# A Sobering Assessment of Small-Molecule Force Field Methods for Low Energy Conformer Predictions

*Ilana Y. Kanal,[1] John A. Keith,[2] Geoffrey R. Hutchison[*,1,2]*

[1]Department of Chemistry, University of Pittsburgh, 219 Parkman Avenue, Pittsburgh, Pennsylvania 15260, United States

[2]Department of Chemical and Petroleum Engineering, Swanson School of Engineering University of Pittsburgh, 804 Benedum Hall, 3700 O'Hara Street, Pittsburgh, PA 15261

ABSTRACT. We have carried out a large scale computational investigation to assess the utility of common small-molecule force fields for computational screening of low energy conformers of typical organic molecules. Using statistical analyses on the energies and relative rankings of up to 250 diverse conformers of 700 different molecular structures, we find that energies from widely-used classical force fields (MMFF94, UFF, and GAFF) show unconditionally poor energy and rank correlation with semiempirical (PM7) and Kohn-Sham density functional theory (DFT) energies calculated at PM7 and DFT optimized geometries. In contrast, semiempirical PM7 calculations show significantly better correlation with DFT calculations and generally better geometries. With these results, we make recommendations to more reliably carry out conformer screening.

**Introduction**. Molecular mechanics (MM) using classical force fields is a highly efficient way to calculate molecular energies and gradients of up to millions of atoms.[1] Given their efficiency, they are also widely used for screening and filtering large numbers of organic molecular structures for atomistic properties, for example for solar materials,[2] computational drug design,[3,4] and/or conformer searching.[5-8] In all cases, the quality of the screening naturally depends on the accuracy of the force fields, and a careful assessment is thus needed to establish their utility in these applications.

The present work focuses on assessing the accuracy of classical small-molecule MM methods frequently used in conformer search applications. Most organic molecules with four or more atoms have some level of conformational flexibility, and even small molecules possess multiple thermally-accessible conformer geometries.[9] Although classical force fields are widely used to identify low energy conformers, recent studies have questioned the reliability of classical force field methods.[10] Kaminský and Jensen have also reported detailed benchmarking studies of conformational energies of amino acids, showing limitations of force fields with fixed charges for biomolecular applications.[8,11] Consequently, in many cases, benchmarks of conformer generation tools are performed, not by considering a low-energy geometry, but by comparing the geometry of an experimental crystal structure against some ensemble (e.g., 50-200+) of conformers.[12,13] Given a reasonable tool, one might guess that generating enough conformers should produce *something* close to the experimental geometry, so finding a method, such as energies, to filter, score or rank conformers is critical. This creates a need for deeper understanding of the limitations of classical force fields across broad chemical applications.

We find several common assumptions are often made to rationalize the use of classical, small-molecule force fields for conformer searches (or other similar applications, such as molecule-protein docking). One assumption is that energy calculations from a classical force field need not be highly accurate to obtain reasonable molecular geometries. A second assumption is that a well-trained force field will be reasonably accurate for molecular structures that fall within the chemical space of the fitted parameterization, even if it performs poorly on species outside of the fitted parameterization. The last assumption is that even though force fields may or may not reliably identify the *lowest* energy conformer, they can be used to locate *low* energy conformers in a reliable fashion. In the present work, we have carried out a comprehensive investigation to assess the validity of each of these assumptions.

**Test Set Selection.** A data set consisting of x-ray crystal structures of 700 small molecules capable of being in multiple conformers was provided to us by Eberjer[12] and were derived from the work of Hawkins *et al*.[13] along with ligands from the Astex Diverse Set.[14] These compounds have been repeatedly used to evaluate the quality of conformer generation.[12,13] Approximately half (320 molecules) consist solely of carbon, hydrogen, nitrogen, and oxygen (CHON) atoms, while the remainder are more complex drug-like compounds and ligands from the Protein Data Bank (PDB).[12] A list of Simplified Molecular Input Line Entry Specification (SMILES)[15] for all 700 molecules can be found in the Supporting Information in Table S1.

**Computational Methods**. While multiple previous works have evaluated the accuracy of different conformer generation tools to produce *geometries* comparable to PDB or other crystal structure geometries, energy ranking is frequently used to remove unlikely geometries. Instead, we generated geometrically diverse conformers using Open Babel[7] for each molecule in the data set. Up to 250 conformers were generated using a genetic algorithm to maximize the heavy-atom root-mean-square displacement (RMSD) between conformers.[16] From the starting geometry of each conformer, conjugate gradient geometry optimizations were performed using Open Babel with the MMFF94,[17-21] UFF,[22] and GAFF[23] classical force fields or with the PM7[24] semiempirical method using OpenMOPAC.[25] Kohn-Sham Density Functional Theory (DFT) electronic energy calculations and geometry optimizations were carried out on subsets of these geometries using ORCA[26] with the B3LYP exchange correlation functional,[27,28] the def2-SVP[29,30] basis set, the RI and RIJCOSX[31] approximations, and the D3BJ[32] dispersion correction scheme. To our knowledge this is the most extensive computational validation set to date for studying low energy molecular conformers.

**Analysis**. Data analysis was performed using Python scripts with the numpy[33] and scipy.stats libraries incorporating the pandas[34] module. We report Spearman correlations that indicate how well two variables correlate as ranked lists (e.g., sorted by energies). A perfect Spearman correlation is +1, and a perfect inverse correlation is –1. Besides Spearman correlations, we also report $R^2$ values, coefficients of intercepts, and slopes of linear correlations for up to 250 conformers for each of the 700 molecules. Data and the Python scripts used to perform the calculations are available at https://github.com/ghutchis/conformer-scoring.

**Results and Discussion**. Energies of each conformer were analyzed with OLS (ordinary least squares) regression as integrated with the scipy and numpy libraries in Python to determine $R^2$ values. Figure 1a illustrates how one $R^2$ value is obtained by calculating the correlation between 250 different conformations optimized using MMFF94 and PM7 for one single molecule ('astex_1l7f'). Note that comparing MMFF94 and PM7 conformer energies consistently results in large scatter and a very low $R^2$ value. Histograms of all $R^2$ values, across the molecular data set are shown in Figures 1(b-d). We find that correlations between classical force fields and semiempirical PM7 are very poor. Spearman rank correlations also demonstrate similarly poor results (Tables S2-S4, Figure S1). Table 1 shows the median and average $R^2$ and Spearman correlation values for all data. Note that for all methods, median and average $R^2$ correlation between force fields and PM7 are a paltry 0.1-0.2.

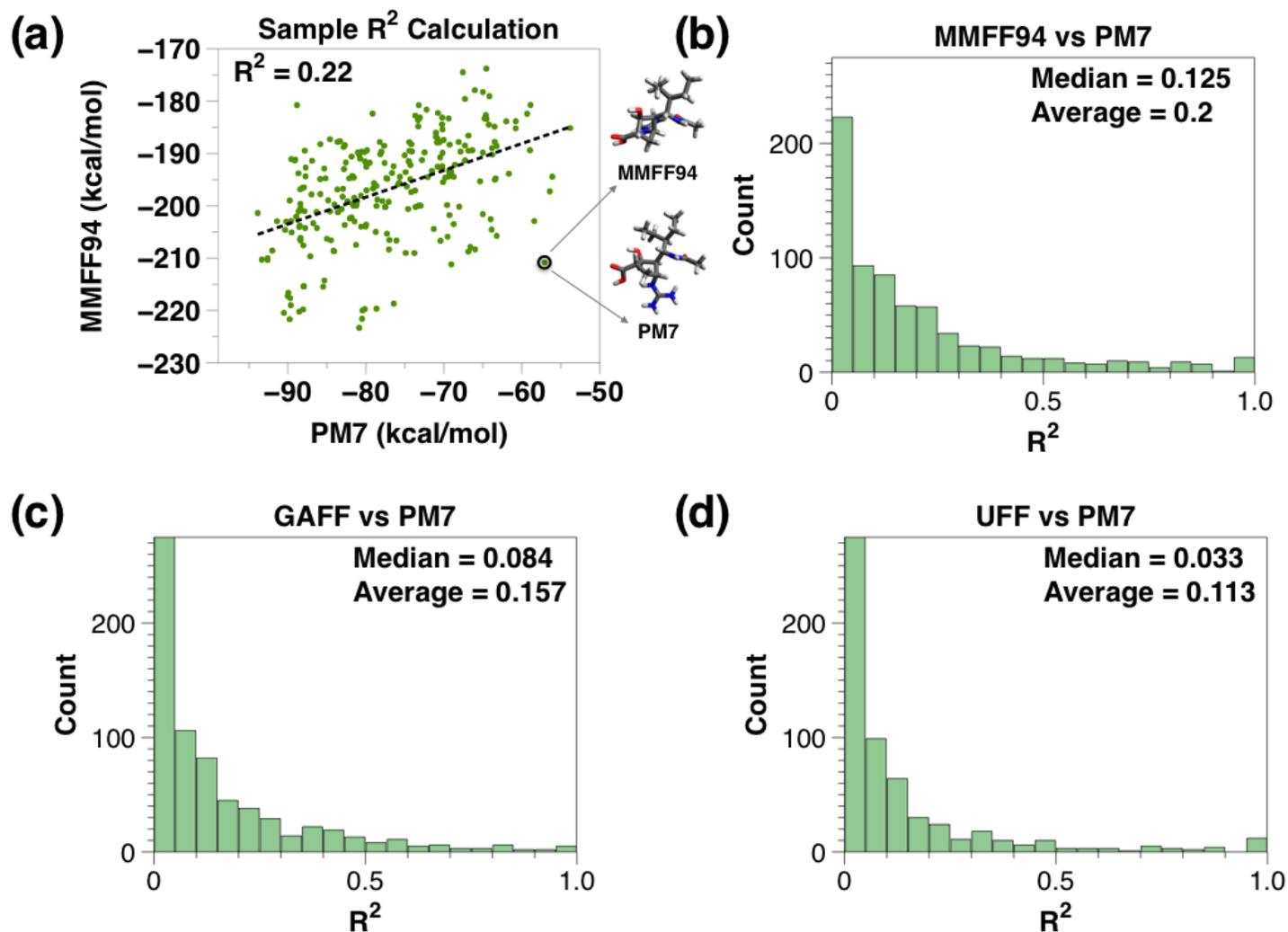

**Figure 1.** (a) 250 data points representing different conformers used to calculate one $R^2$ value for the 'astex_1l7f' molecule. Similar analyses were performed on up to 250 conformers for each of the 700 molecules in our dataset. The circled point on the plot represents a conformer that contributes to the low $R^2$ value by having significantly different MMFF94 and PM7 geometries. (b-d) Histograms of 700 $R^2$ values obtained from the entire data set. MMFF94, GAFF, and UFF all show similarly poor correlation with PM7 methods.

**Table 1.** Median and average $R^2$ values and Spearman correlations for the full data set.

|  | $R^2$ | | Spearman Rank | |
|---|---|---|---|---|
|  | **Median** | **Average** | **Median** | **Average** |
| **UFF vs. PM7** | 0.033 | 0.113 | 0.092 | 0.099 |
| **GAFF vs. PM7** | 0.113 | 0.157 | 0.240 | 0.210 |
| **MMFF94 vs. PM7** | 0.125 | 0.200 | 0.312 | 0.291 |
| **PM7//MMFF94 vs. PM7** | 0.283 | 0.224 | 0.455 | 0.434 |
| **MMFF94 vs. DFT//PM7** | 0.100 | 0.181 | -0.103 | -0.086 |
| **PM7 vs. DFT//PM7** | 0.342 | 0.411 | 0.618 | 0.488 |
| **MMFF94 vs. DFT** | 0.079 | 0.168 | -0.055 | -0.049 |
| **PM7 vs. DFT** | 0.200 | 0.314 | -0.200 | -0.188 |
| **PM7//DFT vs. DFT** | 0.207 | 0.318 | -0.200 | -0.188 |
| **DFT//MMFF94 vs. DFT**[*] | 0.098 | 0.190 | 0.176 | 0.143 |
| **DFT//PM7 vs. DFT** | 0.261 | 0.371 | 0.382 | 0.338 |

* Analyzed across 618 molecules

Although MMFF94 has perceived reliability in generating molecular geometries of organic compounds,[35] these data suggest that all classical force fields have similarly large problems reliably identifying and ranking structurally diverse conformers. *Thus, the assumption that force fields can reliably represent trends in low energy conformers compared to quantum chemistry methods is simply not safe*.

The data above show that MMFF94 demonstrates slightly better correlation with PM7 compared to UFF and GAFF, so we considered PM7 single point energies calculated on MMFF94-optimized geometries (i.e. PM7//MMFF94 calculations). Figures 2(a-b) and Table 1 show that PM7//MMFF94 data has slightly higher median/average $R^2$ values (0.283/0.224) compared to MMFF94 data (0.125/0.200) vs. PM7 data. Figure S2 shows that median/average Spearman rank correlations show a similar trend for MMFF94 (0.312/0.291) and PM7//MMFF94 (0.455/0.434) vs. PM7. Although these results demonstrate slightly improved correlations, the correlation of PM7//MMFF94 with PM7 is still underwhelming, suggesting that MMFF94-optimized geometries are unreliable.

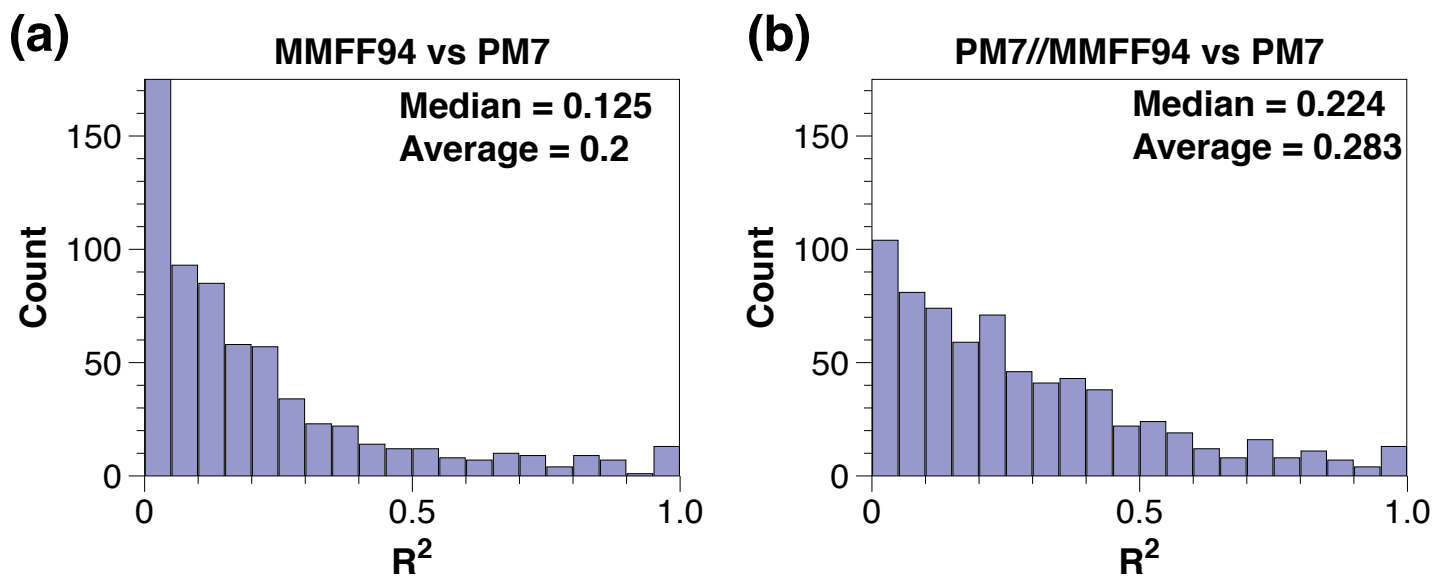

**Figure 2**. Histograms of $R^2$ values obtained using MMFF94 and PM7//MMFF94 data vs. PM7 data. Single point PM7 energy calculations carried out on geometries obtained from classical force fields are only slightly more correlated to the semiempirical optimized geometries.

*Comparison with DFT Single-Point Energies*

We now assess the quality of MMFF94 and PM7 energies and geometries using DFT (B3LYP-D3BJ/def2-SVP calculations). We calculated DFT single point energies (i.e. DFT//PM7 and DFT//MMFF94) for up to ten of the lowest energy conformers from separate PM7 and MMFF94 optimizations on each of the molecules. Although the accuracy of this DFT approach is expected to be deficient compared to more robust electronic structure methods with larger basis sets, it provides a practical representation of a method that should be more reliably accurate than PM7. We also performed geometry optimizations (i.e., DFT//DFT) which will be discussed in the next section. Figure 3 shows histograms of $R^2$ values for MMFF94 and PM7, each vs. DFT//PM7 calculations. (Data found in Tables S6-S8.) The data show that standard MM calculations provide wholly unreliable representations of conformers.

The median/average $R^2$ values are (a) 0.100/0.181 for MMFF94 and (b) 0.342 / 0.411 for PM7 data, each vs. the DFT//PM7 data. Spearman rank correlations show similar results as shown in Figure S3 with median/average values of (a) -0.103/-0.086 for MMFF94, (b) -0.455/0.434 for PM7//MMFF94, and (c) 0.618/0.488 for PM7, each vs. DFT//PM7, as shown in Table 1. Though $R^2$ correlations are not particularly good for PM7 vs. DFT//PM7, the Spearman rank correlations are better, showing that PM7 is more reliable than MMFF94 for ranking conformers.

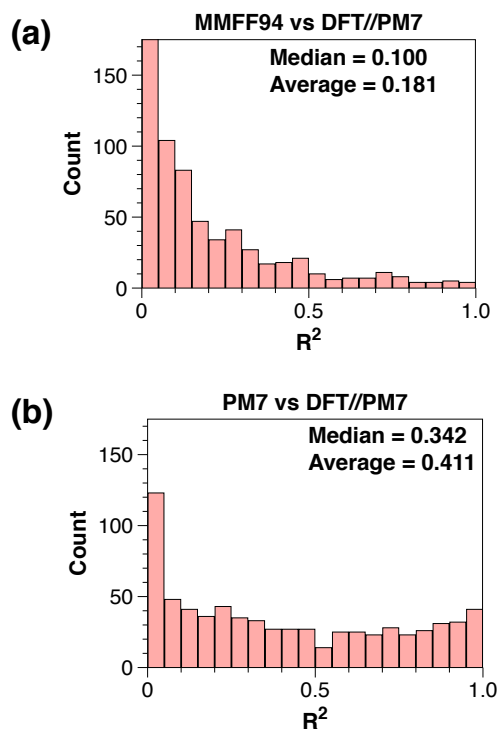

**Figure 3**. Histograms of (a) MMFF94 and (b) PM7, each vs. DFT//PM7 data. Calculations utilizing force fields correlate very poorly with DFT//PM7 data, but PM7 correlates less poorly vs. DFT//PM7.

Figures 4a-b show correlations between DFT//MMFF94 and DFT//PM7 data. Note that Figure 4a shows energies as atomization energies, where the larger number represents a more strongly bound state. The median/average values for $R^2$ values are 0.127/0.223, and Spearman rank correlations (Figure S5) are also similarly poor. Figure 4c shows a histogram of DFT//PM7 – DFT//MMFF94 atomization energies having an average of 1.76 kcal/mol, median of 1.07 kcal/mol and standard deviation of 5.85 kcal/mol. In short, using B3LYP-D3BJ single-point calculations, we find there is frequently a very poor correlation between PM7-optimized and MMFF94-optimized geometries for the same initial conformer. Moreover, Figure 4c indicates that the PM7 geometries are, on average, more stable than corresponding MMFF94 geometries according to the DFT calculations. Still, about a third of the PM7 geometries show worse performance than the corresponding

MMFF94 version. Yet overall, if DFT single-point energies are used, geometries from MMFF94 optimizations are not as reliable as geometries from PM7 optimizations.

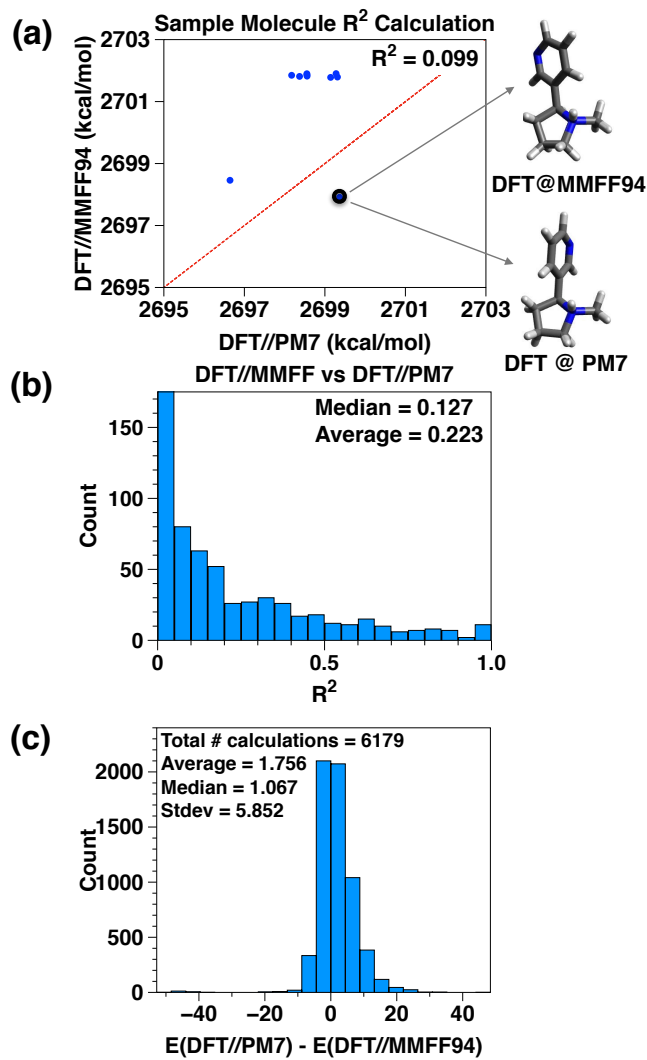

**Figure 4**. (a) 10 data points representing different conformers used to calculate one $R^2$ value for the 'astex_1p2y' molecule. (b) Histograms of 6179 $R^2$ values obtained from a subset of the full data set. The results show poor correlation for DFT//MMFF94 vs. DFT//PM7 data. (c) Histogram of the atomization energy differences E(DFT//PM7) – E(DFT//MMFF94), showing that PM7 optimized geometries are on average lower in energy than MMFF94 geometries within the DFT model.

In short, since the energies of MMFF94 and PM7 are different, the potential energy surfaces *strongly* differ. Even when beginning from the same starting conformer geometry, both methods frequently result in quite different optimized geometries. While none of the methods show strong correlation with one another, the worst correlations with DFT//PM7 data are those that involve classical force fields (Figure S6).

*Comparison with DFT Optimized Geometries*

As noted above, we performed full geometry optimizations using B3LYP-D3BJ and the def2-SVP basis set on up to 10 conformers (i.e., the lowest energy 10 conformers ranked by PM7). This allows comparison between energies and conformer rankings with DFT-quality geometries. Again, average and median $R^2$ correlations and Spearman rank correlations are compiled in Table 1. We note that while neither MMFF94 or PM7 do particularly well at ranking conformers relative to the final DFT optimized energies, $R^2$ correlations for PM7 are somewhat better than MMFF94, although both exhibit median and average Spearman rank correlations *below zero* (i.e., in general, MMFF94 and PM7 tend to rank conformers inversely when compared with DFT optimized rankings).

Since such comparisons rely on different geometries, we also considered the PM7//DFT energies calculated on the DFT optimized geometries. These calculations yield very similar behavior as the PM7-optimized results – with average and median $R^2$ correlations of ~0.2-0.3 with the DFT optimized energies and similar negative Spearman rank correlations.

We also considered DFT single-point energies calculated on MMFF94 and PM7-optimized geometries. The DFT//MMFF94 results correlate very poorly with DFT energies, while the DFT//PM7 results yield average $R^2$ values of 0.37 and average Spearman correlation of 0.34. Thus, the best correlations with full DFT geometry optimizations come, not surprisingly, from DFT single-points on the PM7 optimized geometries, as indicated in Figure 5(a). One reason for the only modest correlation with full optimized geometries comes from the energy strain – Figure 5(b) indicates that on average the PM7 optimized geometry has 6-7 kcal/mol higher atomization energies than the DFT optimized geometry.

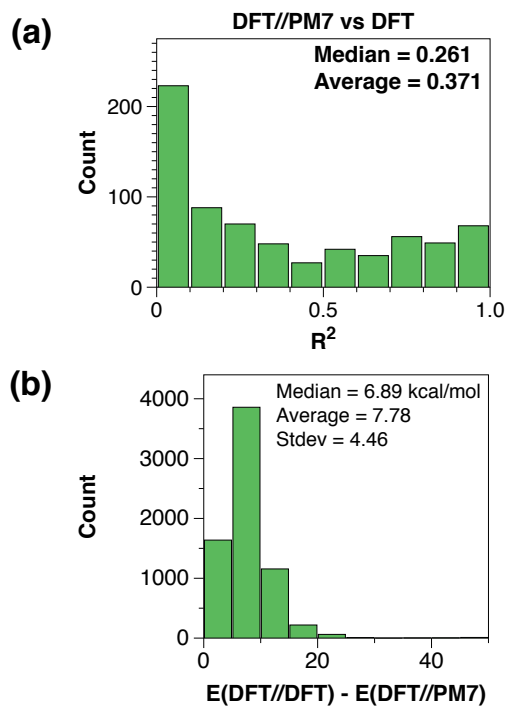

**Figure 5**. (b) Histogram of $R^2$ values between DFT//PM7 and DFT optimized energies and (b) Histogram of the atomization energy differences E(DFT//DFT) – E(DFT//PM7), showing that full geometry optimization with DFT frequently involves ~5-10 kcal/mol relaxation in energy.

*Energetic Ranges: How Many Conformers in an Ensemble?*

Conformer searches aim to identify the most stable conformer or ensemble of conformers. Open source and commercial conformer generation software packages can automate the generation of hundreds or potentially thousands of conformers.[36,37-41] However, as shown above, classical force fields simply do not provide reliable energies or geometries conformer screening.

To identify a practical solution, we determined the fraction of the conformers in our data sets that were within a given energy range of the lowest energy geometry, as computed by a particular method. The number of conformers that were within 1 – 10 kcal/mol at 1 kcal/mol intervals were then counted. Figure 6 summarizes these results and more data are shown in Figures S7-S9.

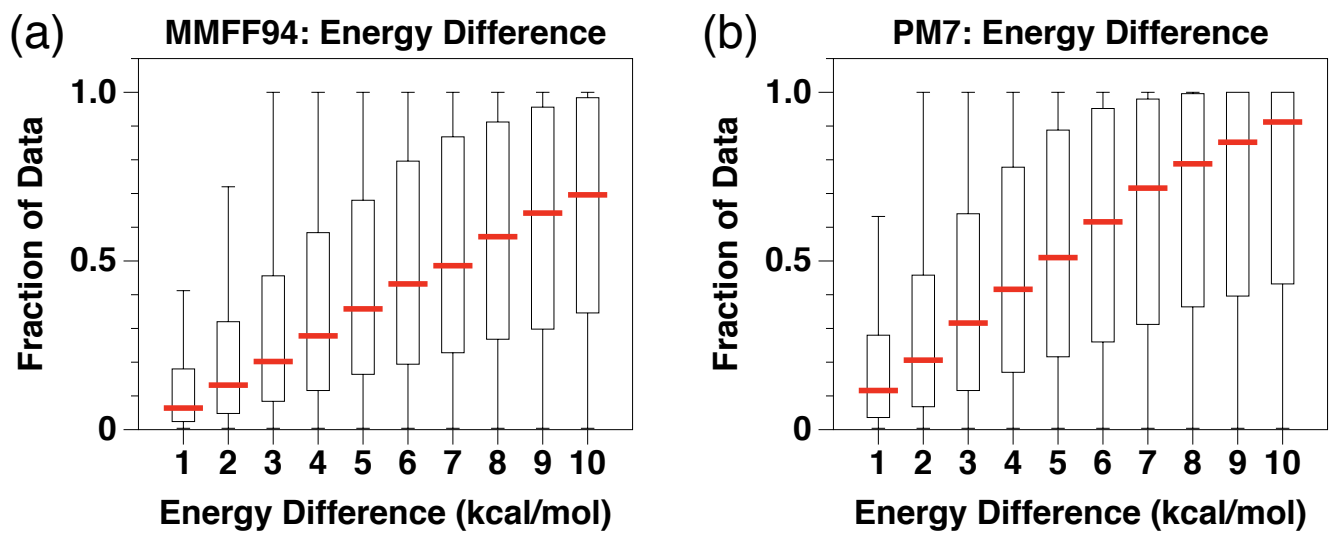

**Figure 6**. Fraction of the data set within energy differences ranging from 1–10 kcal/mol when using (a) MMFF94 and (b) PM7. The red lines represent the median value, the bottom of the square box represents the first quartile, the top of the box represents the third quartile and the endpoints at the top and bottom of the lines represent the maxima and the minima, respectively.

Figure 6a shows that ~6% of the conformers generated using MMFF94 are within 1 kcal/mol of the lowest energy conformer, while ~70% of the conformers generated are within 10 kcal/mol. In the case of PM7 data, ~12% of the conformers are within 1 kcal/mol of the lowest energy conformer and 91% are within 10 kcal/mol. This represents the difficulty in performing conformer rankings, since "chemical accuracy" is typically accepted as ~1 kcal/mol and the validated error in predicted heats of formation for the PM6 and PM7 semiempirical methods are ~8-10 kcal/mol.[24] Consequently, an ideal method for conformer ranking would require predicted thermochemical errors < 0.5 kcal/mol or less.

*Using Force Fields for "Rough" Optimization*

Computationally efficient methods are often used for fast and rough geometry optimization so that fewer optimization steps are needed for further optimizations with quantum methods. Our data indicate that using force fields for rough optimizations is actually inefficient and likely counter-productive. Figure 2 shows that PM7//MMFF94 data poorly correlates with PM7 data, much less DFT. Moreover, the MMFF94 potential energy surface for conformers appears to be very different from that from PM7 (Figure S10). The average PM7 gradient norm when starting from an MMFF94 optimized conformers is ~140 kcal/Å, and the *minimum* gradient norm is ~50 kcal/Å, showing that MMFF94-optimized geometries are often not close to their corresponding PM7-optimized geometries. The average heavy-atom root mean square displacement (RMSD) between MMFF94 and PM7 optimized geometries starting from the same initial state is ~0.6Å (Figure S6).

Moreover, one might imagine the main cause for the large gradients between MMFF94-optimized structures to be bond lengths and angles that might quickly relax. To evaluate this, restricted geometry optimizations were performed on the MMFF94-optimized geometries, using PM7 with frozen dihedral angles. Consequently, bond lengths and angles are relaxed, while retaining any conformational differences between the MMFF94-optimized and PM7-optimized geometry. These new geometries show an average RMSD < 0.2 Å vs. MMFF94, but an average RMSD of >0.5Å vs. PM7-optimized geometries. The main differences between MMF94 and PM7-optimized geometries are **not** the bond lengths and angles, but dihedrals.

Given the large gradient and geometry differences between the two methods, it is thus not a surprise then that MMFF94 and PM7 geometry optimizations result in very different final geometries and very different energy rankings. For this reason, the use of classical MM methods for optimizing molecular structures having multiple torsional degrees of freedom is only advised if the precision and accuracy of the final structures and rankings obtained from the conformer searches is of little or no concern.

*Analysis of Problem Molecules*

Classical force fields are parameterized, and thus it is possible that poor performance reflects a need for improved parameterizations. Some of the molecules in our data set had $R^2$ values uniformly greater than or equal to 0.80 for MMFF94 vs. PM7, MMFF94 vs. DFT//PM7, and PM7 vs DFT//PM7 calculations (Figure 7). In these cases, classical force field parameterizations are doing a respectable job identifying and ranking conformers.

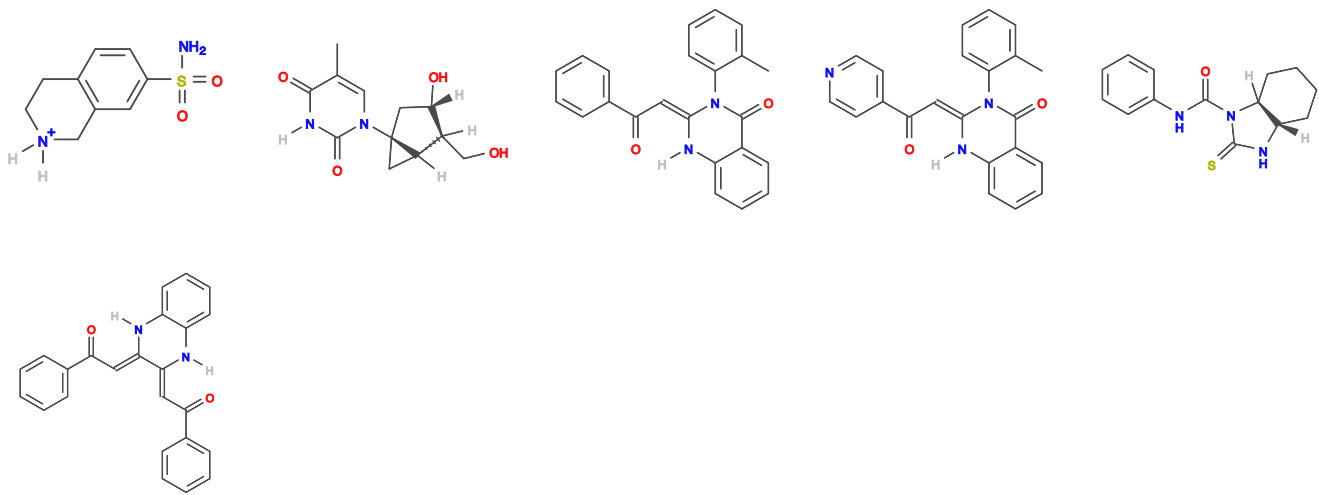

**Figure 7.** Molecules that resulted in $R^2$ values greater than or equal to 0.80 with MMFF94 vs DFT, MMFF94 vs PM7 and PM7 vs. DFT.

There were also cases where molecules had $R^2 \sim 0$, between lower-level methods and higher-level methods (Figure 8). Visual comparison of molecules in Figures 7 and 8 suggests that such molecules have more rotatable torsions and/or contain halides. However, our entire data set was screened using SMARTS patterns for standard functional groups and actually found no statistical evidence of specific functional groups being more present in problem cases than in the well-performing cases. We also note that there were *many* molecules with $R^2$ values near zero. Figures S12-S13 show the 45 molecules with $R^2$ values below 0.05 as calculated from MMFF94 vs. PM7, MMFF94 vs. DFT//PM7, and PM7 vs. DFT//PM7 calculations. Figures S14-S21 show molecules with either low $R^2$ values or greater than 0.80.

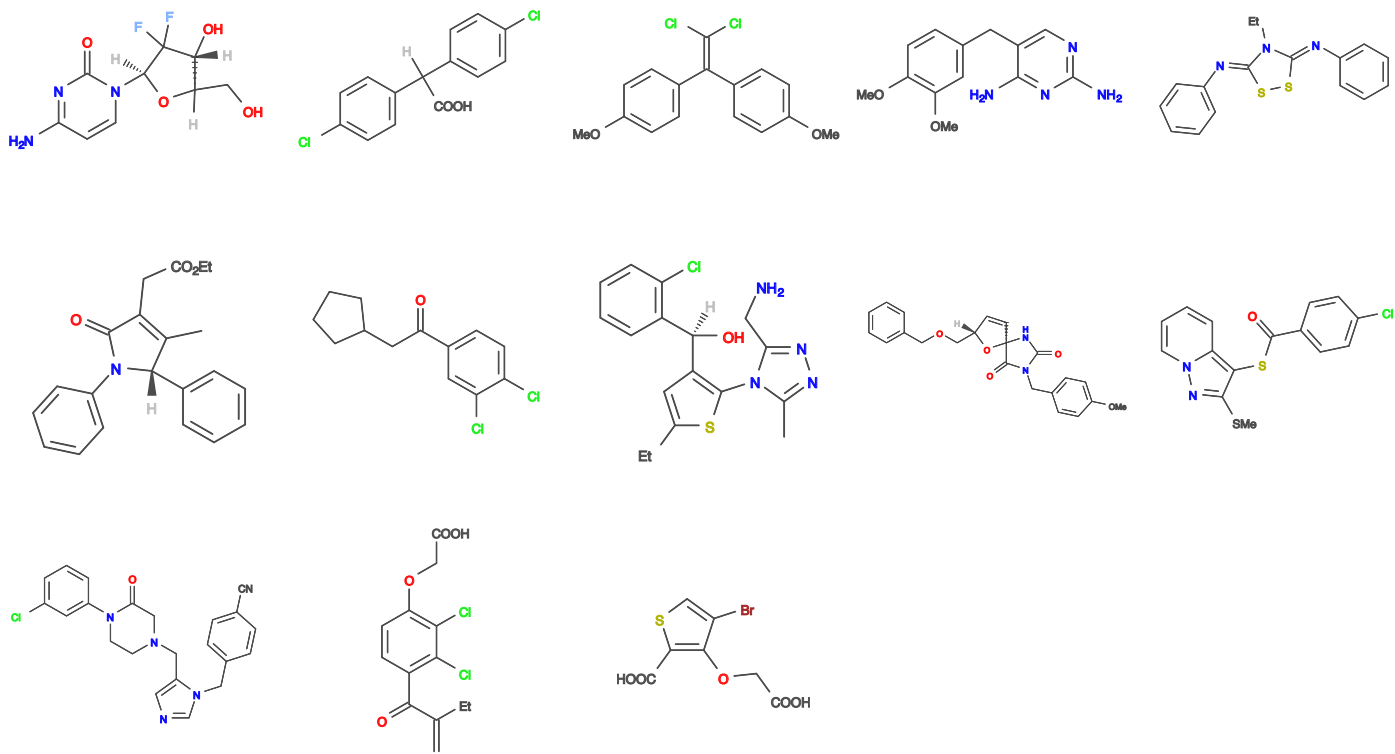

**Figure 8**. Molecules that resulted in poor $R^2$ values across all comparisons of MMFF94 vs DFT//PM7, MMFF94 vs PM7 and PM7 vs. DFT//PM7.

In short, our statistical analysis indicates that the poor performance of MMFF94 (and presumably other classical organic small-molecule force fields) is not simply due to a particular failure in parameterization and that a solution necessarily requires better fitting. Instead, the issue is systematic, and neither the energies nor the optimized geometries of classical force fields should currently be trusted for conformational searching or related applications. The energies, and in turn the potential energy surface produced by general-purpose force fields like MMFF94 in general do not correlate with more accurate quantum chemical methods such as PM7 or even more accurate dispersion-corrected hybrid DFT calculations. Similar investigations of other generic force fields in these applications is warranted, and we will provide all of our dataset free of charge at https://github.com/ghutchis/conformer-scoring for this purpose.

**Conclusions**

We have quantitatively and statistically assessed the accuracy and reliability of classical force fields used in conformer searching applications. Their performances across a large data set of organic molecules shows severe problems that indicate that they are unreliable for conformer searching and/or filtering of low and high energy geometries. Three widely used force fields for general small-molecule chemistry were investigated (i.e., MMFF94, UFF, and GAFF), and all were found to perform similarly poorly. We assess that all are wholly unreliable for conformer screening despite conventional wisdom.

As noted above, conventional assumptions have suggested that even if energies from classical force fields are not entirely accurate, they can produce reasonably high-quality geometries. We actually find that neither classical force field energies nor their geometries seem relatable to data obtained using PM7 or DFT energy calculations. This causes the potential energy surfaces from classical force fields that describe complicated multi-dimensional torsional space to be *very* different from those that would be obtained from quantum methods. Thus, current small molecule force fields should not be trusted to produce accurate potential energy surfaces for large molecules, even in the range of "typical organic compounds."

Moreover, using classical force fields as an initial screen to optimize geometries and/or rank low and high energy geometries makes intuitive sense, but carrying out this procedure with generic classical force fields is likely counterproductive. We find not only large deviation between MMFF94-optimized and PM7-optimized geometries obtained from the same initial structure, but the gradients of the MMFF94 method on a PM7 geometry (and vice versa) are also quite large.

In current applications, we prescribe that regardless of the software used to generate conformer ensembles, one should generate a diverse set of geometries (e.g., using RMSD diversity) and perform geometry optimizations and subsequent energy calculations using the best quantum chemical methods that are tractable. We note that semiempirical methods such as PM7 can be used quite rapidly on modern computing architectures.

That is, since performing an exhaustive evaluation of conformers is time-prohibitive at DFT//DFT level, optimizing multiple conformers with PM7 is tractable, followed by some level of filtering and ranking to compute a subset of single-point DFT energies. We find DFT//PM7 will give you a fairly good correlation with the full DFT//DFT ranking of conformers – especially considering that the differences in energies are often ~1-2 kcal/mol and thus within the method error of B3LYP-D3BJ itself.

We do not mean to suggest that *all* force field methods are unreliable for conformer searching, but we have noted that these problems do not seem to be due to the presence of specific functional groups in some molecules, and thus a need for better parameters. Careful parameterizations, e.g. for biomolecules, and customized force fields derived from quantum chemical methods are certainly useful for specific applications.[42,43] In the short term, we suggest that future parameterizations should attempt to consider more training with non-equilibrium geometries and multiple conformers to ensure that the potential energy surfaces of the force fields better represent quantum chemical methods than they do currently. In the long term, we note that our work highlights an urgent need for methods that can rapidly and reliably screen drug-like organic molecules.

ASSOCIATED CONTENT

**Supporting Information**. SMILES for the 700 molecules in the study, data tables with OLS data values for all calculations reported, histograms of Spearman rank correlations for MMFF94, GAFF, UFF, PM7//MMFF94 compared with PM7, histograms of Spearman rank correlations of MMFF94, MMFF94//PM7 and PM7 compared with DFT//PM7, comparison of $R^2$ values and Spearman Rank correlations of MMFF94//PM7, PM7 and DFT//MMFF94 with DFT//PM7 calculations, histograms the number of conformers for each molecule in the MMFF94 and PM7 data sets that are within 1-10 kcal/ mol. Additional data and Python scripts are available free of charge at https://github.com/ghutchis/conformer-scoring .

AUTHOR INFORMATION

**Author Contributions**

The manuscript was written through contributions of all authors. All authors have given approval to the final version of the manuscript.

ACKNOWLEDGMENT

JAK acknowledges financial support from the Department of Chemical & Petroleum Engineering at the University of Pittsburgh, the Central Research Development Fund, and the R. K. Mellon Foundation. GRH and IYK acknowledge financial support from the NSF (CBET-1404591). We thank Rohith Amruthur, Jeffrey Carr, Yinan Kang, and Yaqun Zhu for help in early stages of this project.

Theory. *Annu. Rev. Phys. Chem.* **2016**, *67* (1), 467–488.